\documentclass{jpp}
\usepackage[utf8]{inputenc}
\usepackage[T1]{fontenc}
\usepackage{amsmath}

\usepackage{color}
\usepackage{graphicx}
\usepackage{upgreek}
\usepackage{combelow}
\usepackage{url}

\usepackage{hyperref}

\hypersetup{
    pdftitle={Advanced surrogate model for electron-scale turbulence in tokamak pedestals},
    pdfauthor={Ionut-Gabriel Farcas, Gabriele Merlo, Frank Jenko},
    pdfkeywords={reduced model ETG turbulence, reduced model ETG, surrogate model ETG turbulence, surrogate model ETG, tokamak pedestal, pedestal, electron-scale turbulence}
}

\definecolor{carminepink}{rgb}{0.92, 0.3, 0.26}
\definecolor{cobalt}{rgb}{0.0, 0.28, 0.67}
\definecolor{cerulean}{rgb}{0.0, 0.48, 0.65}
\definecolor{red}{rgb}{0.92, 0., 0.}

\newcount\Comments
\newcommand{\kibitz}[2]{\ifnum\Comments=1\textcolor{#1}{#2}\fi}

\shorttitle{Advanced surrogate model for electron temperature gradient turbulence}
\shortauthor{I.-G. Farcas, G. Merlo, and F. Jenko}

\title{Advanced surrogate model for electron-scale turbulence in tokamak pedestals}

\author{Ionu\cb{t}-Gabriel Farca\cb{s}\aff{1,2}
  \corresp{\email{ionut.farcas@austin.utexas.edu}},
  Gabriele Merlo\aff{3,4,1}
 \and Frank Jenko\aff{3,4,1}}

\affiliation{\aff{1}Oden Institute for Computational Engineering and Sciences, The University of Texas at Austin, TX 78712, USA
\aff{2}Department of Mathematics, Virginia Tech, VA 24061, USA
\aff{3}Max Planck Institute for Plasma Physics, Boltzmannstr.~2, 85748 Garching, Germany
\aff{4}Institute for Fusion Studies, The University of Texas at Austin, TX 78712, USA}

\begin{document}

\maketitle

\begin{abstract}

We derive an advanced surrogate model for predicting turbulent transport at the edge of tokamaks driven by electron temperature gradient (ETG) modes. 
Our derivation is based on a recently developed sensitivity-driven sparse grid interpolation approach for uncertainty quantification and sensitivity analysis at scale, which informs the set of parameters that define the surrogate model as a scaling law. 
Our model reveals that ETG-driven electron heat flux is influenced by the safety factor $q$, electron beta $\beta_e$, and normalized electron Debye length $\lambda_D$, in addition to well-established parameters such as the electron temperature and density gradients.
To assess the trustworthiness of our model's predictions beyond training, we compute prediction intervals using bootstrapping.
The surrogate model's predictive power is tested across a wide range of parameter values, including within-distribution testing parameters (to verify our model) as well as out-of-bounds and out-of-distribution testing (to validate the proposed model). 
Overall, validation efforts show that our model competes well with, or can even outperform, existing scaling laws in predicting ETG-driven transport.
\end{abstract}

\section{Introduction} \label{sec:intro}

Turbulent transport is known to determine the energy confinement time of fusion devices.
Quantifying, predicting, and controlling this turbulent transport is a prerequisite for designing optimized fusion power plants and is therefore considered a key open problem in fusion research.
While high-fidelity numerical simulations based on first principles are crucial for understanding the complex mechanisms behind turbulent transport, they are often computationally too expensive for routine use in tasks like optimization or uncertainty quantification that involve large ensembles of simulations. 
Constructing computationally cheap yet reliable surrogate models is therefore highly desirable in practice.

In~\citet{FMJ22}, we proposed a sensitivity-driven dimension-adaptive sparse grid framework for uncertainty quantification (UQ) and sensitivity analysis (SA) at scale, which was applied to the problem of turbulent transport in the pedestal region of a tokamak driven by electron temperature gradient (ETG) modes. 
The sensitivity-driven approach intrinsically provides an interpolation-based surrogate model, which turned out to provide accurate predictions for the scenario studied in~\citet{FMJ22} for testing parameters within the interpolation bounds. 
However, since polynomial extrapolation is, in general, ill posed and unstable~\citep{Tr12}, we will not use this model for predictions outside the interpolation bounds used for its construction.
In addition, this interpolation model does not incorporate a quantification of prediction uncertainty, which is crucial for assessing the trustworthiness of surrogate transport models for any given input parameters.
The present paper builds on~\citet{FMJ22} and leverages the aforementioned sensitivity-driven approach to derive a generic and parsimonious surrogate transport model for ETG-driven turbulent fluxes that (i) depends on physically interpretable parameters, (ii) generalizes beyond training, and, importantly, (iii) incorporates a quantification of prediction uncertainty.  

Electron temperature gradient turbulence has been the subject of theoretical and numerical investigations for more than two decades, from the seminal works by~\citet{Je00, Dorland, JD02} focused on core plasmas to more recent studies~\citep{told, jenko_09, Idomura, Hatch_2015, Hatch_2017,  Hassan_2022,Chapman-Oplopoiou_2022,Li_2024, Pa20, Pa22, St22, Le23, BCS23, BCS24} indicating its role in regulating transport in the pedestal.
Given its importance, several recent papers such as~\cite{Chapman-Oplopoiou_2022, Guttenfelder_2021, Ha22, Hatch_2024} have formulated surrogate models as well as simple algebraic expressions for ETG fluxes in the pedestal. 

Here, we propose a scaling law for the ETG-driven electron heat flux as a function of the safety factor $q$, the electron beta $\beta_e$, and the normalized electron Debye length $\lambda_D$, in addition to well-established parameters such as electron temperature and density gradients. 
The exponents appearing in our scaling law are obtained by performing a simple regression fit using the existing high-fidelity, nonlinear simulation results from~\cite{FMJ22}. 
To the best of our knowledge, the dependence on $q$ and $\beta_e$ is new and describes the effect of the magnetic geometry on ETG modes. 
Furthermore, we incorporate a quantification of uncertainty in the predictions issued by the proposed surrogate model, a critical requirement in data-driven modelling, essential for evaluating the predictive performance for arbitrary input parameters. 
To this end, we compute prediction intervals via bootstrapping (we refer the reader to \cite{EB86} for a review of bootstrapping methods), which offers a distribution-free and reliable method to account for data variability and model uncertainty. 
We test the prediction capabilities of the proposed surrogate model across a wide range of parameter values. 
The model is verified against $32$ within-distribution testing parameters and validated against $61$ out-of-distribution and $40$ out-of-bounds testing parameters. 
We also compare it with similar scaling laws available in the literature, obtaining similar or more accurate predictions.

The remainder of this paper is organized as follows.
Section~\ref{sec:background} summarizes the baseline simulation parameters used to derive the proposed surrogate model.
Section~\ref{sec:findings} details the steps used to derive our surrogate model including how to incorporate a quantification of uncertainty in its predictions by computing prediction intervals via bootstrapping.
Section~\ref{sec:results} presents our results.
We assess the prediction capabilities of the proposed model and compute prediction intervals via boostrapping over a wide range of parameter values, from within-distribution testing data to out-of-bounds and out-of-distribution testing parameters.
Section~\ref{sec:scaling_dep_beyond_grad} investigates the origin of the parametric dependencies appearing in the proposed model, in particular $q$ and $\beta_e$. 
Finally, some overall conclusions are presented in Section~\ref{sec:conclusions}.
The code and data to reproduce our results are publicly available at \url{https://github.com/ionutfarcas/surrogate_model_etg_pedestal}.

\section{Baseline simulation parameters} \label{sec:background}

The present paper builds on the UQ and SA study performed in~\cite{FMJ22} and derives a generic and parsimonious surrogate transport model for ETG turbulence.
In the following, we present the aspects relevant for this work and refer the reader to~\cite{FMJ22} for further details. 

The scenario under consideration models DIII-D conditions similar to~\cite{Hassan_2022,walker} and is representative of typical pedestal conditions, where the large gradients can drive a substantial electron heat flux via ETG turbulence; we refer the reader to~\cite{Hassan_2022} for a detailed description of the experimental conditions.
We consider that the plasma behaviour is fully specified by the following eight local parameters, $\{n_e, T_e, \omega_{n_e}, \omega_{T_e}, q, \hat{s}, \tau, Z_\mathrm{eff} \}$. Here, $n_e$ is the electron density, and $T_e$ is the electron temperature, $\omega _{n_e}=a/L_{n_e} = 1/n_e \textrm{d} n_e / \textrm{d} \rho_{\textrm{tor}} $ and $\omega _{T_e} = a/L_{T_e} = 1/T_e \textrm{d} T_e / \textrm{d} \rho_{\textrm{tor}}$ are the respective normalized (with respect to the minor radius $a$) gradients, where $\rho_{\textrm{tor}}$ denotes the denotes the square root of the normalized toroidal magnetic ﬂux, $q$ is the safety factor, $\hat{s}=r/qdq/dr$ is the magnetic shear ($r$ denotes the minor radius defined below), $\tau=Z_{\rm eff}T_e/T_i$, and $Z_\mathrm{eff}$ is the effective ion charge retained in the collisions operator (here, a linearized Landau--Boltzmann operator).
The eight parameters are modelled as independent uniform random variables with symmetric bounds around their nominal values; for more details, see Appendix~\ref{app:params}. 
Moreover, the ions are assumed to be adiabatic, and electromagnetic effects, computed consistently with the values of the electron temperature and density, are retained.

The magnetic geometry is specified according to a generalized Miller parametrization~\citep{Candy_2009}. 
We employed $64$ Fourier harmonics to achieve a sufficiently accurate representation of the flux-surface at $\rho_{tor}=0.95$, near the pedestal top. 
Using a generalized Miller parametrization instead of a magnetohydrodynamics (MHD) equilibrium allows varying independently (and self-consistently) $q$ and $\hat{s}$. 
Note that the parametrization of the magnetic geometry is typically affected by uncertainties as well and should therefore be incorporated into the underlying set of uncertain inputs. 
This, however, would increase the number of uncertain inputs drastically, making the UQ and SA study significantly more challenging; we leave such an extended analysis for future work.

The simulations in~\cite{FMJ22} were carried out using the plasma micro-turbulence simulation code \textsc{Gene}~\citep{Je00} in the flux-tube limit using a box with $n_{k_x}\times n_{k_y}\times n_{z}\times n_{v_\|}\times n_{\mu}=256\times24\times168\times32\times8$ degrees of freedom in the five-dimensional position--velocity space.
This grid was shown to be sufficiently fine by performing convergence tests conducted for several input configurations including the four pairs comprising the gradients extrema,  corresponding to the highest and lowest electron heat flow values within the considered parameter range. 
We note that several works in the literature~\citep{Chapman-Oplopoiou_2022, Guttenfelder_2021, walker} observed that the computed values of ETG fluxes depend sensitively on the resolution $n_{z}$ in the parallel direction.
A resolution comprising $n_z = 168$ points was sufficient in our case because we leveraged the input parameter \texttt{edge\_opt} in \textsc{Gene}, which allows use of a lower resolution by redistributing the grid points in the parallel direction.
Subsequent simulations in this paper utilize this grid resolution.

\section{Parsimonious surrogate model with prediction uncertainty quantification}
\label{sec:findings}

The sensitivity-driven approach in~\cite{FMJ22} enabled an efficient UQ and SA in the ETG scenario under consideration at a cost of only $57$ nonlinear \textsc{Gene} simulations.
This low number of simulations was due to the fact that this approach is adaptive, with a refinement indicator based on sensitivity information about the uncertain inputs.
In this way, the adaptive procedure preferentially refines the directions corresponding to important input parameters and interactions thereof.
Our goal here is to derive an interpretable and predictive surrogate model for ETG turbulent transport with quantified prediction uncertainty.
This will be achieved by leveraging information provided by the sensitivity-driven approach, the readily available $57$ simulation results, and bootstrapping for computing prediction intervals. 

A byproduct of the sensitivity-driven approach is that it also intrinsically provides an interpolation-based polynomial surrogate of the output of interest in terms of the uncertain inputs (which can be trivially mapped to a spectral projection basis, a Legendre basis in our case).
The sparse interpolation surrogate in~\cite{FMJ22} approximated the power crossing the flux-surface in MW due to ETG turbulence in terms of $\{n_e, T_e, \omega_{n_e}, \omega_{T_e}, q, \hat{s}, \tau, Z_\mathrm{eff} \}$.
As it turns out, this surrogate is accurate for within-distribution testing points, that is, eight-dimensional input parameters falling within the considered uncertainty bounds.
For example, the work in~\cite{FMJ22} showed that the mean-squared approximation error at $N = 32$ pseudo-random testing parameters was in $\mathcal{O}(10^{-4})$.
The goal of the present paper is to obtain a general surrogate model that is predictive for parameter values beyond the considered uncertainty bounds and, importantly, also incorporates a measure of prediction uncertainty.
Since it is well-established that polynomial extrapolation is generally ill posed and unstable~\citep{Tr12}, we will not exploit the sparse grid surrogate but target a more compact scaling law that relates the turbulent flux to key plasma parameters. 
The results provided by the sensitivity-driven approach will guide us in choosing which plasma parameters should define the target scaling law.

Prior to describing how the proposed surrogate model is obtained, we address two important details. 
The initial point of clarification pertains to units and normalizations.
A surrogate model in S.I.~units would be preferable because these units are the most general. 
However, we cannot construct such a model here without redoing all gyrokinetic simulations from~\cite{FMJ22} since the set of uncertain inputs does not include a macroscopic length $L_{\rm ref}$ nor a magnetic field strength $B_{\rm ref}$.
Even if these two parameters would turn out to be unimportant, such an assessment cannot be made \emph{a priori} without redoing the UQ and SA study.
We instead opt for a model for the electron heat flux in gyroBohm (GB) units, $Q_{\rm GB}=n_eT_e^{5/2}m_i^{1/2}/(eB_{\rm ref}L_{\rm ref})^2$, which provide a natural normalization for our set-up.
This implies that we must map the original $57$ simulation results in SI~units to their associated heat fluxes in GB units, $Q_e/Q_{\mathrm{GB}}$. 
Since the flux-surface area is constant, no particular issue arises from removing it from the existing $57$ simulation results. 
However, the GB units depend on two out of the eight uncertain inputs, namely $n_e$ and $T_e$.
Because of this, there is no guarantee that the adaptive procedure in the sensitivity-driven approach would produce the same $57$ simulation results in GB units as in the original S.I.~units.
Nevertheless, the existing simulation results can be reused irrespective of the units and, as we will show in our results, using GB units does not, in fact, impact the accuracy of the obtained surrogate model.

The second point of clarification concerns the definition of the normalizing quantities. 
It is crucial to precisely define reference quantities like $L_{\rm ref}$ and $B_{\rm ref}$ to prevent inaccurate model predictions that may not align with experimental measurements, for example.
The same applies to input parameters such as density and temperature gradients. 
A quantity that impacts the values of these parameters is the selection of the radial coordinate.
In the following, we assume that the GB units are defined in terms of the magnetic field on the axis, $B_0$, and the effective minor radius, $a=\sqrt{\Phi_{\rm LCFS}/B_0}$, where the toroidal flux value at the last closed flux-surface (LCFS), $\Phi_{\rm LCFS}$, represents the length parameter. 
This implies that the radial coordinate in the definition of the gradients is $\rho_{tor}=\sqrt{\Phi/\Phi_{\rm LCFS}}$, where $\Phi$ denotes the toroidal flux.
It is nevertheless important to keep in mind that these choices directly impact the surrogate model that we will derive next.
This implies, for example, that we cannot rule out the possibility that using different definitions, such as alternative radial coordinates, may result in a simpler surrogate~\cite{Hatch_2024}.

\begin{figure}
    \centering
    \includegraphics[width=1.0\textwidth]{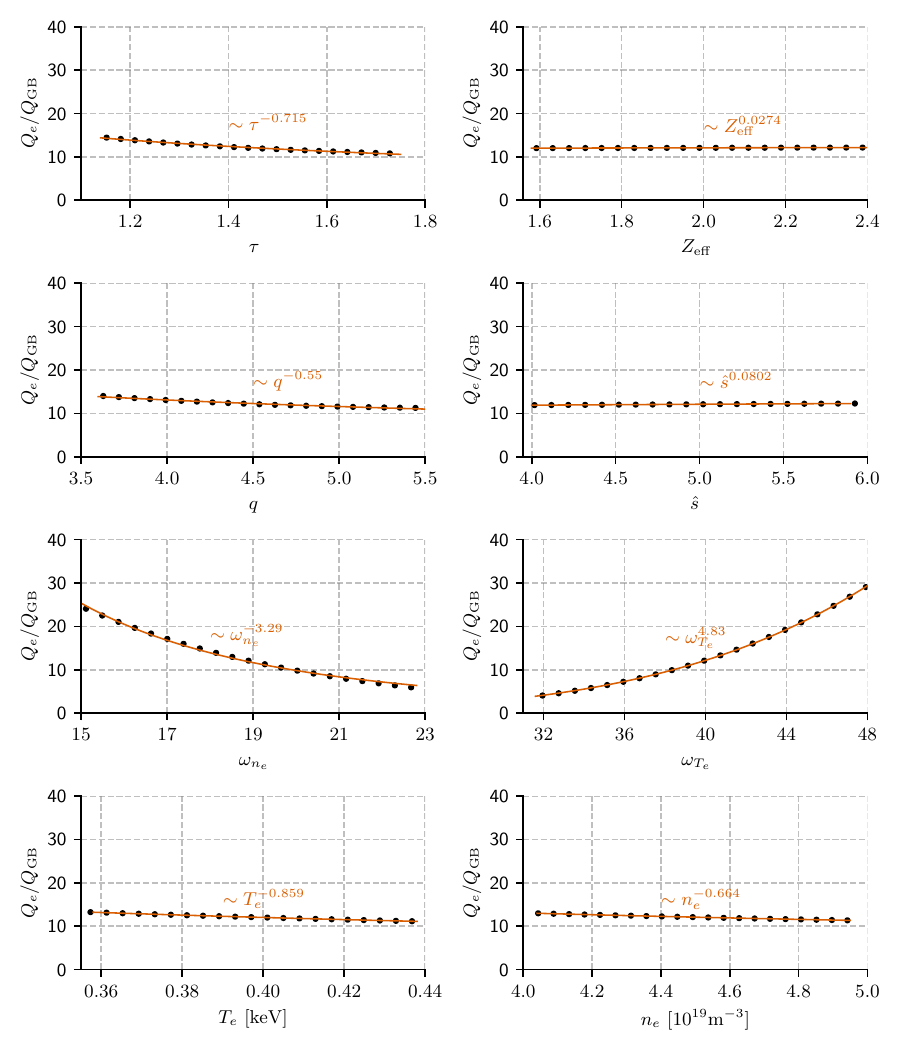}
    \caption{Dependence of the electron heat flux on each of the eight considered parameters obtained using the sparse grid surrogate model. The remaining seven parameters are fixed to their respective nominal values. We also estimate via regression the rates at which the flux varies with the eight inputs.}
    \label{fig:Q_1D_dependencies_nominal}
\end{figure}

Given the choice of normalizing quantities and radial coordinate specified above, we must decide next which input parameters should specify our surrogate model.
We seek a set of parameters that are both important, in the sense that variations in these parameters lead to non-negligible variations in the electron heat flux, as well as physically interpretable.
To this end, we leverage the information provided by the sensitivity-driven approach.
More specifically, we use the fact that this approach provides an accurate sparse grid interpolation  surrogate, which, in turn, can be employed to assess the dependence of the heat flux in terms of subsets of uncertain inputs. 
Using its representation in the Legendre basis, this surrogate, in GB units, amounts to a multi-variate polynomial of third degree comprising $47$ terms; its expression is provided in~Eq.~\eqref{eq:Q_SGI_8D_full} in Appendix~\ref{app:SGI}.
Figure~\ref{fig:Q_1D_dependencies_nominal} plots the one-dimensional dependencies of $Q_e/Q_{\mathrm{GB}}$ in terms of all eight uncertain inputs, obtained by fixing the seven other parameters to their respective nominal values.
We also perform regression fits that give the rates at which the heat flux varies with each parameter. 
We remark that these dependencies do not account for parameter interactions (such interactions would appear in higher-dimensional dependencies).
Moreover, some rates may change if the remaining seven parameters (e.g., the two gradients) were fixed to other values, such as their extrema.

Nonetheless, the obtained results indicate that the two gradients lead to the most significant dependencies, which is in line with what was expected.  
In contrast, the dependencies in terms of $\hat{s}$ and $Z_{\rm eff}$ are clearly negligible, implying that $\hat{s}$ and $Z_{\rm eff}$ can be ignored in our surrogate model.
We observe small but non-negligible dependencies due to $\{\tau, q, n_e, T_e\}$. 
Thus, these results suggest a dependence on $\{\omega_{T_e}, \omega_{n_e}, q, \tau, n_e, T_e\}$.
Naively using these input parameters is not desirable since $n_e$ and $T_e$ are not easily interpretable. 
Instead, by examining the gyrokinetic equations, we identify interpretable parameters depending on $n_e$ or $T_e$ that may enter our model.
These are the collisionality (measured by an appropriate collision frequency $\nu_c$), Debye screening (described by the normalized electron Debye length $\lambda_D$; in the following we will use $\lambda_{D}=\lambda_{De}/\rho_e$ normalized with the electron Larmor radius $\rho_e$, the natural microscopic length scale for our problem), and plasma beta $\beta_e=403\times10^{-5} n_e T_e/B_0$ (here the electron density $n_e$ is measured in $10^{19}/m^3$, $T_e$ in keV, and $B_0$ in T), which affects the magnetic geometry and causes magnetic fluctuations.
However, since we cannot perform a simple change of variables from $\{n_e, T_e\}$ to $\{\lambda_{D}, \nu_c, \beta_e\}$, we instead perform additional simulations to identify which subset of $\{\lambda_{D}, \nu_c, \beta_e\}$ should enter our surrogate model.
We note that considering $\{\lambda_{D}, \nu_c, \beta_e\}$ instead of $\{n_e, T_e\}$ allows us to more clearly pinpoint possible physical effects, and leads to a more interpretable and intuitive surrogate model. 

We perform additional scans in which collisions, Debye shielding, or electromagnetic effects are individually switched off.
For a comprehensive perspective, we perform these scans for both left and right uniform bounds of $\{\omega_{T_e}, \omega_{n_e}, q, \tau, n_e, T_e\}$ used in the sensitivity-driven approach (corresponding to their respective smallest and largest values; see table~\ref{tab:parameters} in Appendix~\ref{app:params}). 
Figure \ref{fig:hists} plots the results. 
We observe a clear dependence on $\beta_e$ and a weaker but not negligible one on $\lambda_{D}$. 
Collisions, in contrast, do not lead to any significant variations in the heat fluxes which implies that $\nu_c$ is unimportant. 
Based on these results, our surrogate model should depend on $\{\omega_{T_e}, \omega_{n_e}, q, \tau, \beta_e, \lambda_{D}\}$. 
Lastly, since it is well-established that ETG transport is a threshold process with respect to $\eta_e = \omega_{T_e}/\omega_{n_e}$, with finite fluxes only when $\eta_e\gtrsim1$ as established in~\cite{Jenko_pop_2001, Je00}, we use $\eta_e$ instead of $\omega_{n_e}$.
We therefore conclude that the input parameters that define our surrogate model are $\{\omega_{T_e}, \eta_e, q, \tau, \beta_e, \lambda_{D}\}$.

\begin{figure}
\centering
\includegraphics[width=\textwidth]{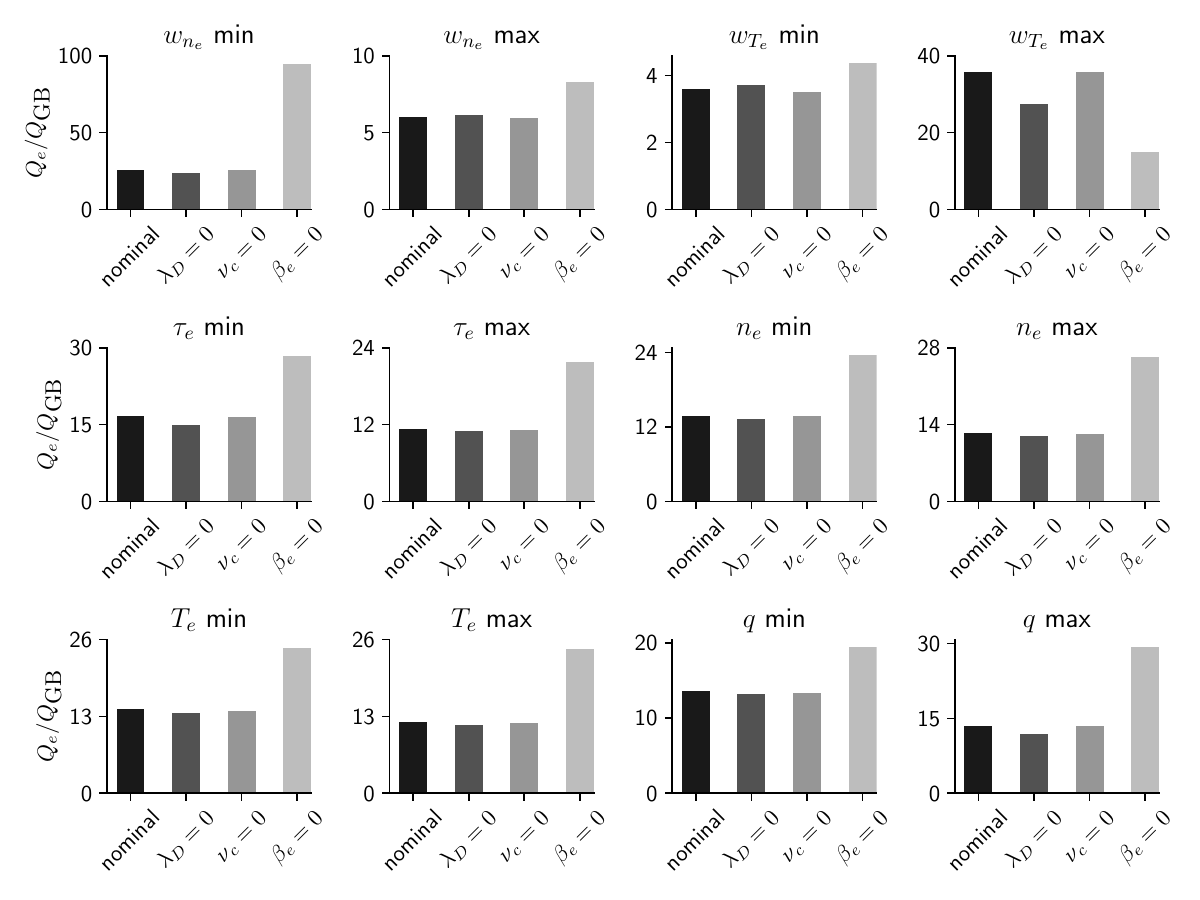}
\caption{Dependence of the electron heat flux on various physical effects. Each panel compares the nominal heat flux with the one obtained when, individually, Debye shielding ($\lambda_{D}$), collisions ($\nu_c$), or electromagnetic effects ($\beta_e$) are excluded. In each case, all other plasma parameters except the one indicated in the title are kept to their respective nominal values.}
\label{fig:hists}
\end{figure}

We seek a surrogate model as a scaling law of the form
\begin{equation} \label{eq:our_ETG_red_model_unfitted}
    Q_e/Q_{\mathrm{GB}} = c_0 \sqrt{\frac{m_e}{m_i}}\,  \omega_{T_e}^{p_1} (\eta_e - 1)^{p_2} \tau_e^{p_3} q^{p_4} \beta_e^{p_5} \lambda_{D}^{p_6},
\end{equation}
where $\{c_0, p_1, p_2, p_3, p_4, p_5, p_6\} \in \mathbb{R}^7$.
To determine these coefficients, we perform a data fit using the existing $57$ nonlinear \textsc{Gene} simulations from our UQ and SA study from~\cite{FMJ22}, where we map the original inputs $\{\omega_{T_e}, \omega_{n_e}, q, \tau, n_e, T_e\}$ to $\{\omega_{T_e}, \eta_e, q, \tau, \beta_e, \lambda_{D}\}$ and the corresponding outputs in SI units to heat fluxes in GB units. 
We note that fitting the power law~Eq.~\eqref{eq:our_ETG_red_model_unfitted} directly is not straightforward as the target function is nonlinear.
Even more, a poor choice of the loss function or minimization approach can lead to a biased model that overfits and hence generalizes poorly.
We take advantage of the fact that we want to fit a scaling law with strictly positive parameters and perform a standard regression fit in logarithmic coordinates.
We obtain:
\begin{equation} \label{eq:our_ETG_reduced_model}
    Q_e/Q_{\mathrm{GB}} = 0.038 \sqrt{\frac{m_e}{m_i}}\, \omega_{T_e}^{1.40} (\eta_e - 1)^{1.79} \tau_e^{-0.76} q^{-0.51} \beta_e^{-0.87} \lambda_{D}^{-0.51}.
\end{equation}

To assess the validity of the predefined threshold $\eta_{e, 0} = 1$ in the proposed model, we perform an additional experiment in which $\eta_{e, 0}$ is a free parameter.
That is, we consider the surrogate model
\begin{equation*}
    Q_e/Q_{\mathrm{GB}} = c_0^{\prime} \sqrt{\frac{m_e}{m_i}}\, \omega_{T_e}^{p_1^{\prime}} (\eta_e - \eta_{e, 0})^{p_2^{\prime}} \tau_e^{p_3^{\prime}} q^{p_4^{\prime}} \beta_e^{p_5^{\prime}} \lambda_{D}^{p_6^{\prime}},
\end{equation*}
where $\{c_0^{\prime}, p_1^{\prime}, \eta_{e, 0}, p_2^{\prime}, p_3^{\prime}, p_4^{\prime}, p_5^{\prime}, p_6^{\prime}\} \in \mathbb{R}^8$.
We determine these coefficients analogously to~\eqref{eq:our_ETG_red_model_unfitted} and obtain
\begin{equation} \label{eq:our_ETG_reduced_model_with_threshold}
    Q_e/Q_{\mathrm{GB}} = 0.048 \sqrt{\frac{m_e}{m_i}} \,\omega_{T_e}^{1.40} (\eta_e - 1.03)^{1.70} \tau_e^{-0.76} q^{-0.52} \beta_e^{-0.85} \lambda_{D}^{-0.41}.
\end{equation}
Model~\eqref{eq:our_ETG_reduced_model_with_threshold} is similar to the model in~\eqref{eq:our_ETG_reduced_model} and in fact both models yield almost identical predictions for all datasets considered in Section~\ref{sec:results}.
This fit therefore reveals that the threshold $\eta_{e, 0} = 1$ is valid in our context.
We note, however, that we cannot disregard the fact that $\eta_{e, 0}$ may have a different value.
Assessing this would require performing simulations with $\eta_{e}$ values close to the threshold, which, in our context would again necessitate redoing the sparse grid study since the minimum $\eta_{e}$ value for the $57$ sparse grid points was $\eta_{e, \mathrm{min}} = 1.4$.
We leave such an analysis for future research.

In practice, the deterministic predictions from the model in~\eqref{eq:our_ETG_reduced_model} are not enough. 
Generally, in order to enhance the reliability of surrogate models, particularly those derived from data fitting, it is important for their predictions to include an element of prediction uncertainty.
To tackle this issue, we calculate prediction intervals using bootstrapping.
Bootstrapping offers a reliable method for establishing prediction intervals without relying on restrictive assumptions regarding the distribution of data and errors.
To compute prediction intervals, we resample the $57$ input--output pairs $M \in \mathbb{N}$ times with replacement and fit a model for each of the $M$ resampled pairs via regression as described above.
We then compute the target predictions for each fitted model, which results in an ensemble of $M$ predictions.
Lastly, this ensemble is used to calculate prediction intervals.

\section{Predictions using the novel surrogate model} \label{sec:results}

We now assess the prediction capabilities of the novel surrogate model given in Eq.~\eqref{eq:our_ETG_reduced_model} and compute prediction intervals using bootstrapping.
For a comprehensive perspective, we perform three sets of experiments comprising $133$ testing points in total across a wide range of values, including two sets that go beyond the training data.

We will compare our model with the most accurate surrogate proposed in~\cite{Ha22}, obtained via symbolic regression using a database of $N = 61$ ETG simulations comprising discharges from the JET, DIII-D, ASDEX Upgrade, and C-MOD tokamaks:
\begin{equation} \label{eq:model_hatch_et_al_2022}
    Q_{e}/Q_{\mathrm{GB}} = \sqrt{\frac{m_e}{m_i}}\, \omega_{T_e}(1.44 + 0.50\,\eta_e^4).
\end{equation}
To the best of our knowledge, this represents the most comprehensive such database currently available in the literature.
The more recent work in~\cite{Hatch_2024} refined the formulas from~\cite{Ha22} and proposed new, richer models that provide more accurate predictions for the database.
We therefore additionally consider the model given in Eq.~(1) in~\cite{Hatch_2024}:
\begin{equation} \label{eq:model_hatch_et_al_2024}
    Q^{\prime}_{e}/Q_{\mathrm{GB}} = 0.019 \sqrt{\frac{m_e}{m_i}}\, \left(\omega_{T_e}^{\prime}\right)^2 (\eta_e - 1)\eta_e^{1.57} \tau_e^{-0.5} \left(\lambda_D^{\prime}\right)^{-0.4},
\end{equation}
where the quantities marked with a prime symbol are defined differently than in the present paper.
More specifically, $\omega_{T_e}^{\prime}$ denotes the temperature gradient taken with respect to the normalized poloidal flux, $\psi$, and $\lambda_D^{\prime}$ denotes the Debye length normalized to $\rho_s = \sqrt{T_e/m_i} m_i / {e B}$.
In addition, the evaluation of the heat flux uses the effective flux-surface area $A = 2\pi R 2 \pi a$, where $R$ denotes the major radius and $a$ the minor radius (for our case, $R=1.68$ m and $a=0.77$ m) instead of the true flux-surface area $V^\prime$.
As noted in Section~\ref{sec:findings}, the conversion factors between radial coordinates has some dependence on $q$ and $\beta_e$.
We therefore cannot exclude the possibility that a model formulated for the radial coordinate $r$ (as in this paper) has different $q$ and $\beta_e$ dependence than a model formulated for the radial coordinate $\psi$ as in~\eqref{eq:model_hatch_et_al_2024}.

To use the model in~\eqref{eq:model_hatch_et_al_2024} in our context, we must (i) convert our temperature gradient and Debye length to be compatible with the units used in~\eqref{eq:model_hatch_et_al_2024} and (ii) convert the ensuing heat flux predictions to be compatible with our definition by accounting for the different flux-surface area.
The conversion of the Debye length is straightforward, namely $\lambda^{\prime}_D = \lambda_D \sqrt{m_e/m_i}$.
The conversion for the temperature gradient and heat flux is explained in detail in Appendix~\ref{app:coversion_to_Hatch_2024}.
As noted there, a full conversion of the temperature gradient is not possible because it requires knowledge of a global MHD equilibrium. 
Nonetheless, we employ the model in~\eqref{eq:model_hatch_et_al_2024} in our experiments both for a more comprehensive perspective and also because this model is richer than the model in~\eqref{eq:model_hatch_et_al_2022}. 
For more details about model~\eqref{eq:model_hatch_et_al_2024}, we refer the reader to~\cite{Hatch_2024}.
We furthermore note that other works such as~\cite{Chapman-Oplopoiou_2022} and~\cite{Guttenfelder_2021} provide algebraic trends of ETG fluxes, which, however, do not represent general surrogate ETG models as their expressions depend on undetermined constants.
The work in~\cite{Hatch_2024} showed that the models proposed therein, including~\eqref{eq:model_hatch_et_al_2024}, yield more accurate predictions than the aforementioned algebraic expressions with constants fitted using the database in~\cite{Ha22}.

In all our experiments, we use $M = 1,000$ bootstrapping samples and compute $95\%$ prediction intervals.
Furthermore, to assess the prediction accuracy of the deterministic predictions issued by our model (i.e., without taking into account the prediction intervals) in a manner consistent with common practice in the literature, we will utilize the following error metric that compares a set of $N$ reference electron heat fluxes (hereby denoted by $Q_{\mathrm{e, ref}}/Q_{\mathrm{GB}}$) and corresponding (deterministic) approximations obtained via a surrogate model (denoted by $Q_{\mathrm{e, approx}}/Q_{\mathrm{GB}}$), $\{Q_{\mathrm{e, ref}; i}/Q_{\mathrm{GB}},  Q_{\mathrm{e, approx}; i}/Q_{\mathrm{GB}}\}_{i=1}^N$: 
\begin{equation} \label{eq:error_measure}
    \varepsilon(Q_{\mathrm{e, ref}}/Q_{\mathrm{GB}}, Q_{\mathrm{e, approx}}/Q_{\mathrm{GB}}) = \sqrt{\frac{1}{N} \sum_{i=1}^N \frac{(Q_{\mathrm{e, ref}; i}/Q_{\mathrm{GB}} - Q_{\mathrm{e, approx}; i}/Q_{\mathrm{GB}})^2}{(Q_{\mathrm{e, ref}; i}/Q_{\mathrm{GB}} + Q_{\mathrm{e, approx}; i}/Q_{\mathrm{GB}})^2}}
\end{equation}
This error metric was also adopted in~\cite{Ha22} and~\cite{Hatch_2024} and it equally penalizes extreme cases, that is, $Q_{\mathrm{e, approx}} \ll Q_{\mathrm{e, ref}}$ and $Q_{\mathrm{e, approx}} \gg Q_{\mathrm{e, ref}}$.

\subsection{Within-distribution testing parameters}

We initially investigate whether the proposed regression-based scaling law~\eqref{eq:our_ETG_reduced_model} exhibits any significant decrease in accuracy compared with the sensitivity-driven sparse grid surrogate model obtained from~\citet{FMJ22}, which was shown to be accurate for within-distribution testing data.
For this purpose, we utilize the $N = 32$ testing samples from~\cite{FMJ22}.
We map the original input parameters $\{\omega_{T_e}, \omega_{n_e}, q, \tau, n_e, T_e\}$ to $\{\omega_{T_e}, \eta_e, q, \tau, \beta_e, \lambda_{D}\}$ and the corresponding outputs to heat fluxes in GB units.
In addition, we also assess the predictions provided by models~\eqref{eq:model_hatch_et_al_2022} and~\eqref{eq:model_hatch_et_al_2024}.

\begin{figure}
\centering
\includegraphics[width=1.0\textwidth]{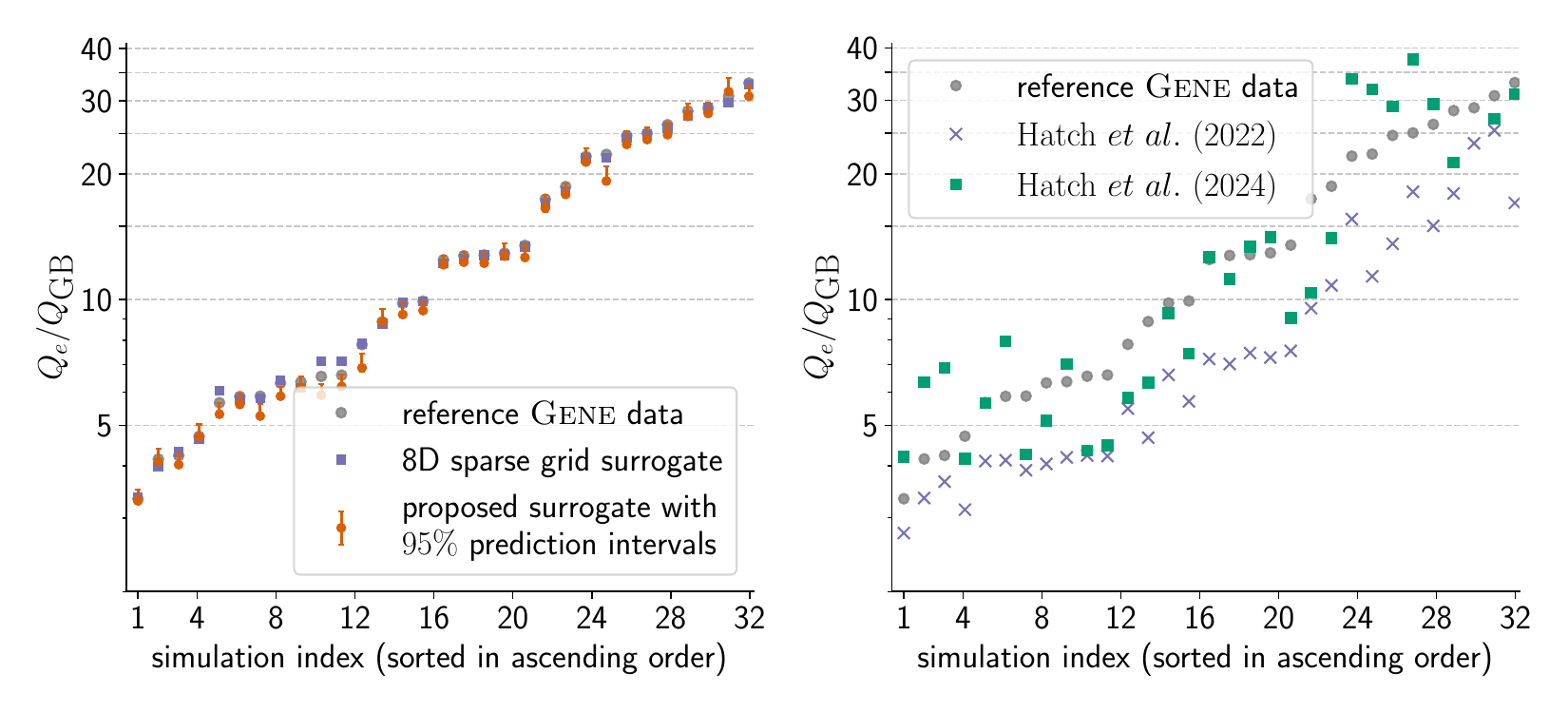}
\caption{Left: comparison between the proposed surrogate model (with $95\%$ prediction intervals) and the more complex sparse grid surrogate model depending on all eight uncertain inputs from~\cite{FMJ22} in GB units using $N = 32$ within-distribution testing data points. Right: the corresponding predictions using models~\eqref{eq:model_hatch_et_al_2022} and~\eqref{eq:model_hatch_et_al_2024}. To simplify visualization, we reordered the fluxes in ascending order relative to the reference values.}
\label{fig:model_verification}
\end{figure}

Figure~\ref{fig:model_verification} shows the results.
To simplify visualization, we reordered the results in ascending order relative to the reference results.
The left plot shows the predictions obtained using our surrogate model and the sparse grid model based on the work in~\citet{FMJ22}.
Both surrogate models provide accurate predictions that are close to each other.
In addition, the $95\%$ prediction intervals, shown for each heat flux predicted by our model, are small.
The corresponding deterministic errors are $\varepsilon = 0.0145$ for the sparse grid model and $\varepsilon = 0.0292$ for the proposed surrogate model.
We can therefore conclude that the proposed regression-based scaling law~\eqref{eq:our_ETG_reduced_model} provides accurate predictions with small prediction intervals that do not deteriorate the accuracy of the more complex sparse grid surrogate for within-distribution testing data.
The right plot shows the predictions obtained with models~\eqref{eq:model_hatch_et_al_2022} and~\eqref{eq:model_hatch_et_al_2024}.
The model in~\eqref{eq:model_hatch_et_al_2022} slightly underpredicts the reference fluxes, but nevertheless yields an error $\varepsilon = 0.2248$.
The richer model~\eqref{eq:model_hatch_et_al_2024} improves these predictions, decreasing the error to $\varepsilon = 0.1561$.

\subsection{Testing the model using data beyond training}
Next, we evaluate the predictive performance of the proposed model on data that fall outside the input distribution. 
We examine two datasets with parameters that exceed the training boundaries. 
The first dataset consists of $61$ out-of-distribution testing points from the database in~\cite{Ha22}, while the second dataset includes $40$ out-of-bounds testing points. 
These experiments serve as validation tests for our model.

\subsubsection{Out-of-distribution testing parameters}
We test the proposed model using the database of $N = 61$ ETG simulations from~\cite{Ha22}.
Given that the $61$ entries pertain to four distinct machines and different configurations, and none of the input pairs $\{\omega_{T_e}, \omega_{n_e}, \tau, \beta_e$, $\lambda_{D}\}$ from the database fall within the uniform bounds of our inputs, it is reasonable to consider these $61$ data points as out-of-distribution.

\begin{figure}
\centering
\includegraphics[width=0.7\textwidth]{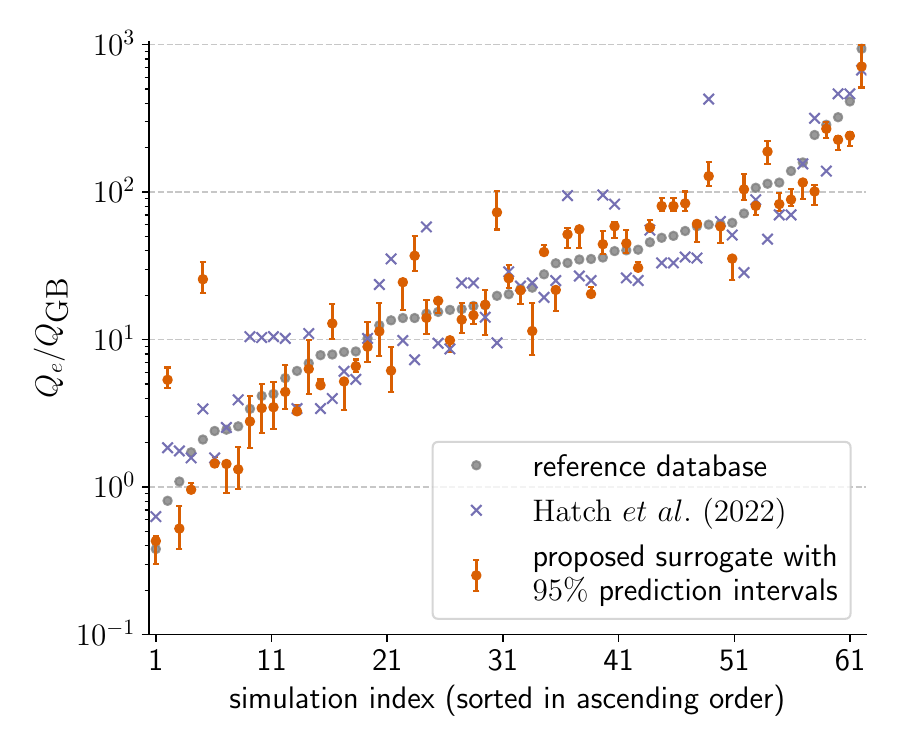}
\caption{Comparison between the proposed surrogate model (with $95\%$ prediction intervals) and the surrogate model~\eqref{eq:model_hatch_et_al_2022} at the $N = 61$ data points from the database in~\cite{Ha22}. These points represent out-of-distribution testing data for our model. The refined surrogate model~\eqref{eq:model_hatch_et_al_2024} provides more accurate predictions with error $\varepsilon = 0.15$. To simplify visualization, we reordered the fluxes in ascending order relative to the reference values.}
\label{fig:model_validation_database}
\end{figure}

Figure~\ref{fig:model_validation_database} plots the results.
This experiment highlights the benefits of including prediction intervals: these intervals not only indicate the level of uncertainty in our predictions but also enhance our understanding of the predictive capabilities of our model.
Specifically, $17$ reference fluxes are within the $95\%$ prediction intervals of our model forecasts, and most of the other reference fluxes are close to our prediction intervals with the exception of a handful of outliers.
There are five outliers for which the respective relative error exceeds $100\%$ (corresponding to indices $16, 17, 19, 20$, and $46$ in the database in~\cite{Ha22}).
Correspondingly, the model in~\eqref{eq:model_hatch_et_al_2022} yields $10$ such outliers.
The value of the deterministic error~\eqref{eq:error_measure} for our model is $\varepsilon = 0.2604$, which is slightly smaller than the error of the surrogate model given by~\eqref{eq:model_hatch_et_al_2022}, $\varepsilon = 0.2860$, which was trained using the entire database.
We note, however, that model~\eqref{eq:model_hatch_et_al_2024} (the corresponding results are not plotted in Figure~\ref{fig:model_validation_database} because the necessary conversion factors are not available in the database in~\cite{Ha22}) provides more accurate predictions (in the units used in~\cite{Hatch_2024}), reducing the error by nearly $50\%$ to $\varepsilon = 0.15$. 

\begin{figure}
\centering
\includegraphics[width=0.7\textwidth]{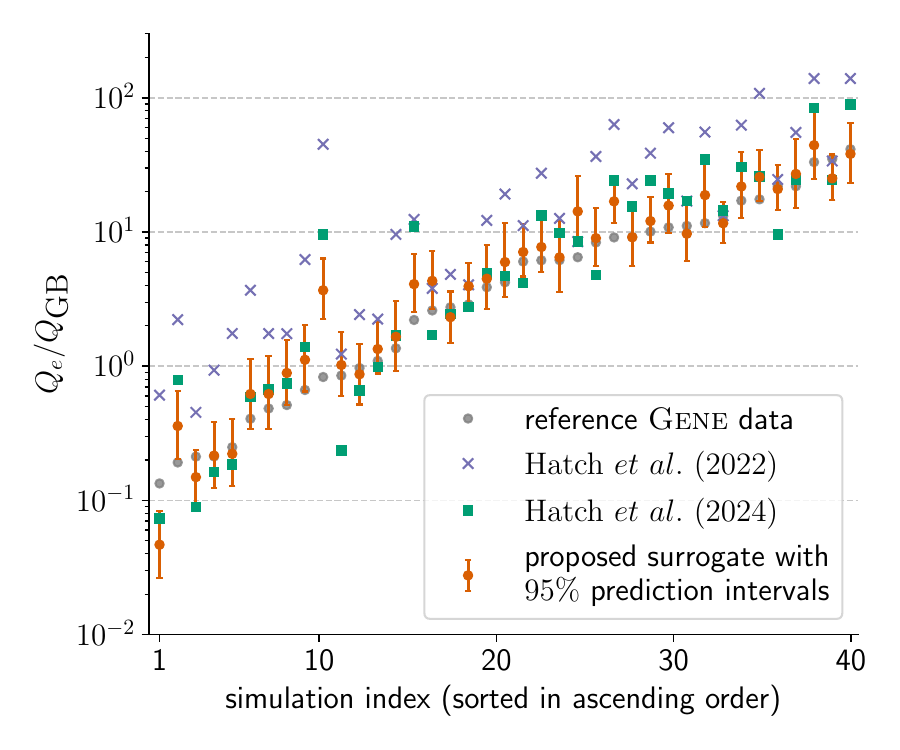}
\caption{Comparison between the predictions obtained using the proposed surrogate model (plus their corresponding $95\%$ prediction intervals) and surrogates~\eqref{eq:model_hatch_et_al_2022} and~\eqref{eq:model_hatch_et_al_2024} at $N = 40$ out-of-bounds testing points. To simplify visualization, we reordered the fluxes in ascending order relative to the reference values.}
\label{fig:model_validation_out_of_distr}
\end{figure}

\subsubsection{Out-of-bounds testing parameters} \label{subsubsec:out_of_bounds_validation}

Lastly, for a more comprehensive assessment of our proposed surrogate model, we test its prediction accuracy using out-of-bounds testing data.
For this experiment, we generate $N = 40$ uniform testing samples that fall outside of the training bounds given in table~\ref{tab:parameters} in Appendix~\ref{app:params}.
The values of the $N = 40$ input parameters as well as the corresponding heat fluxes in GB units are listed in table~\ref{tab:out_of_distr_data} in Appendix~\ref{app:out_ouf_bounds_testing_data}.

Figure~\ref{fig:model_validation_out_of_distr} plots the results.
Our surrogate model closely matches the reference results.
In fact, $35$ out of the $40$ reference flux values are within or very close to our prediction intervals.
The corresponding deterministic error amounts to $\varepsilon = 0.2074$.
In contrast, the surrogate given in Eq.~\eqref{eq:model_hatch_et_al_2022} produces fairly inaccurate predictions that overestimate the reference heat fluxes, resulting in a large error $\varepsilon = 0.5451$.
We attribute these results to the fact that the model~\eqref{eq:model_hatch_et_al_2022} is not rich enough for these testing points as it depends on only two parameters, $\omega_{T_e}$ and $\eta_e$.  
In contrast, the refined surrogate model~\eqref{eq:model_hatch_et_al_2024} produces more accurate predictions that decrease the error down to $\varepsilon=0.3280$.

\section{On the scaling parameter dependencies}~\label{sec:scaling_dep_beyond_grad}
We examine in more detail the dependence of the new scaling law on the parameters $\{\tau, \lambda_{D}, \beta_e, q\}$.
With this aim, we conduct supplemental simulations in which each of the four parameters is varied individually, with all others held constant at their nominal values. 

\begin{figure}
\centering
\includegraphics[width=0.6\textwidth]{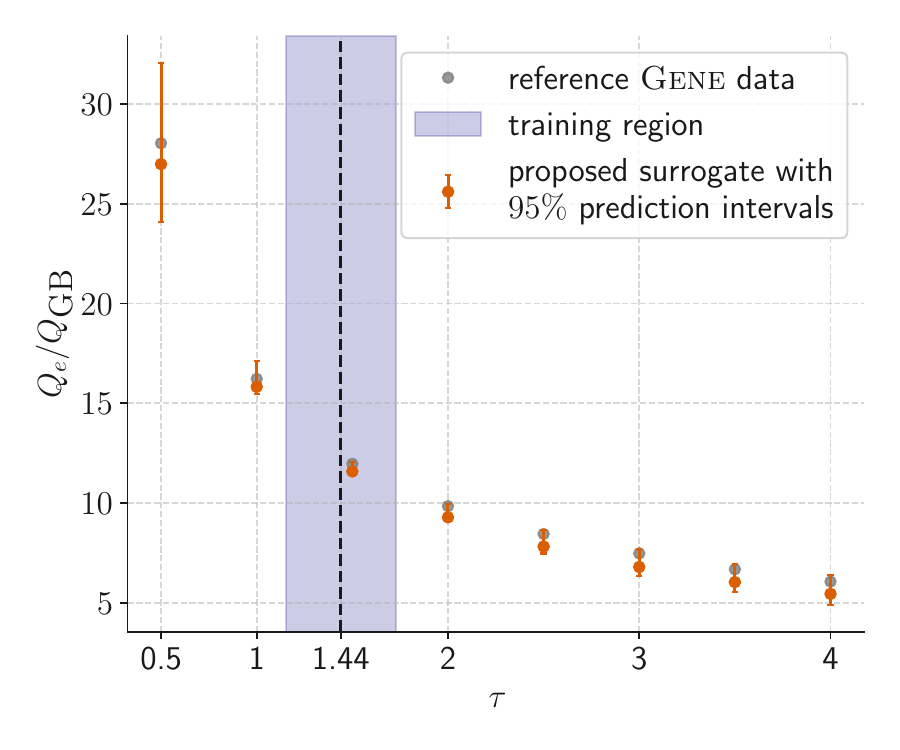}
\caption{Dependence of the electron heat flux on $\tau$ values that exceed the training bounds. We compare the reference \textsc{Gene} data with the predictions obtained using our surrogate model (plus their corresponding $95\%$ prediction intervals).} 
\label{fig:dep_tau}
\end{figure}

Figure \ref{fig:dep_tau} shows the dependence of the electron heat flux on $\tau$. 
We explore a substantially broader spectrum of values for this parameter than those used for the construction of the surrogate model, while ensuring these values are realistic and accounting for a possible large impurity content. 
Our results indicate a reduction in the heat flux, which is consistent with previous findings. 
For instance, the study by~\cite{Jenko_pop_2001} found that the linear dynamics of ETG modes shares similarities with ion temperature gradient (ITG) modes, albeit with the electrons and ions being interchanged. 
Consequently, an increase in the electron-to-ion temperature ratio, $\tau$, is likely to exert a stabilizing influence on ETG-driven fluxes. 
We compare the reference \textsc{Gene} data with the predictions obtained using our surrogate model in Eq.~\eqref{eq:our_ETG_reduced_model} (plus their corresponding $95\%$ prediction intervals) by fixing all parameters except for $\tau$ to their respective nominal values in our surrogate.
Our predictions in Figure \ref{fig:dep_tau} not only reflect this relationship but do so with high accuracy.
In fact, all reference \textsc{Gene} values fall within the $95\%$ prediction intervals.

\begin{figure}
\centering
\includegraphics[width=\textwidth]{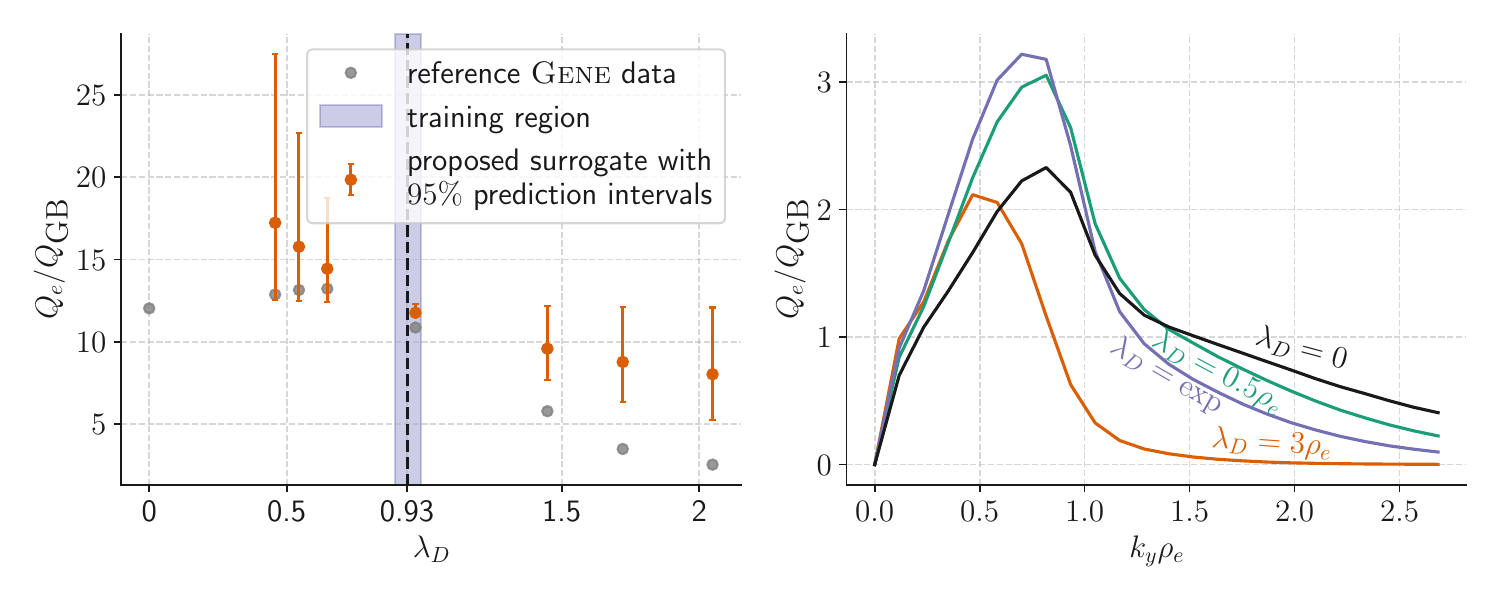}
\caption{ Left: dependence of the electron heat flux on $\lambda_{D} \geq 0$ values that exceed the training bounds. The reference \textsc{Gene} data are compared with the predictions obtained via our surrogate (plus their $95\%$ prediction intervals) for $\lambda_{D} > 0$. Right: flux spectra for different values of $\lambda_{D}$.}
\label{fig:dep_ld}
\end{figure}

We proceed with examining the impact of varying the normalized electron Debye length $\lambda_{D}$. 
The corresponding results are plotted in Figure \ref{fig:dep_ld}. 
The left plot compares the reference \textsc{Gene} data with the predictions obtained via our surrogate model (plus their corresponding $95\%$ prediction intervals) for $\lambda_{D} > 0$ by setting all parameters except for $\lambda_{D}$ to their respective nominal values.
As $\lambda_{D}$ exceeds unity, we observe a decrease in the electron heat flux, corroborating our expectations. 
For instance, this is consistent with the observed stabilization of the high-$k_y$ modes with increasing Debye length, as illustrated in the right plot in Figure \ref{fig:dep_ld}. 
The nominal value $\lambda_{D} = 0.93$ lies within the range where Debye shielding starts to exert a significant influence on transport. 
We remark that the deviation between \textsc{Gene} and the surrogate predictions for large values of  $\lambda_{D}$ was expected since the surrogate model was constructed without exploring this region.
This, however, does not limit the scope of our model since typical plasma parameters of current and future pedestals do not fall within this region.

\begin{figure}
\centering
\includegraphics[width=0.6\textwidth]{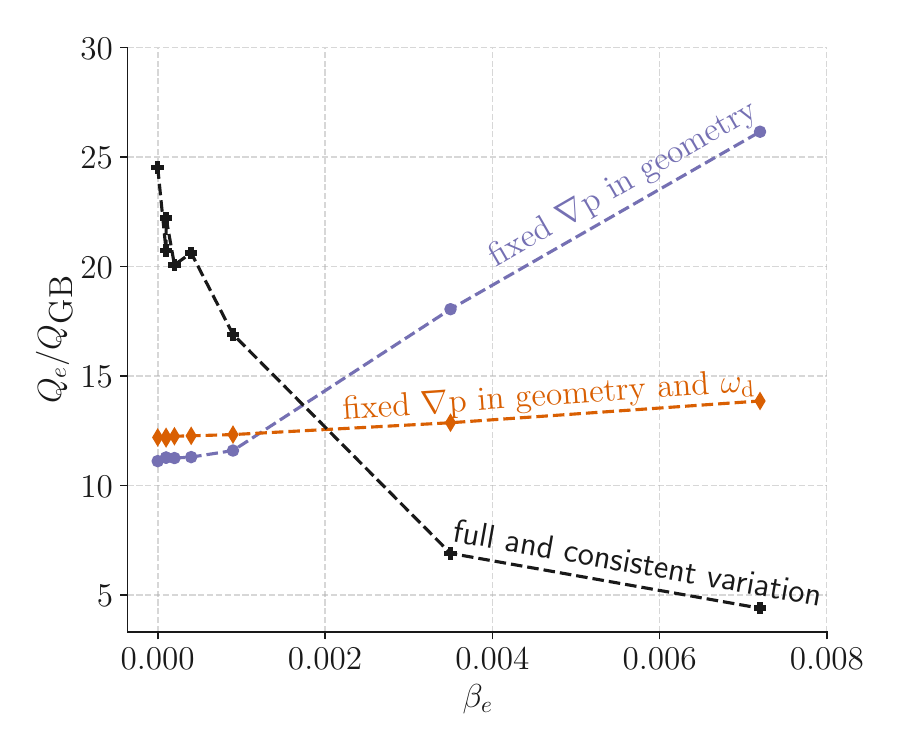}
\caption{Dependence of the electron heat flux on $\beta_e$. The black line plots the simulation results where all terms in the gyrokinetic equations are modified consistently. The orange line shows the results in which the pressure gradient $\nabla p$ is held constant when evaluating particle drifts and the magnetic equilibrium. The purple line plots the simulation results where $\nabla p$ is only kept constant when determining the magnetic equilibrium.}
\label{fig:dep_beta}
\end{figure}

Investigating the effects of the remaining two parameters, $\beta_e$ and $q$, presents a more considerable challenge due to their multifaceted role in the gyrokinetic equations. Specifically, $\beta_e$ influences not only the field equations but also the particle dynamics through the pressure gradient $\nabla p$, which contributes to the drift velocity. 
Additionally, within the context of magnetic geometry, $\beta_e$ plays a critical in determining the Shafranov shift, again via the pressure gradient. 
When working with normalized parameters, we obtain
\begin{equation}
\nabla p = -\beta_e\sum_j\frac{n_j}{n_e} \frac{T_j}{T_e}\left(\omega_{T,j}+\omega_{n,j}\right),
\end{equation}
where the sum runs over all plasma species. 
Consequently, we conduct parameter scans in which we vary $\beta_e$ and simultaneously compute all the affected terms. 
We then compare the results with those from scenarios in which we maintain a constant pressure gradient to isolate and evaluate the impact of the magnetic geometry on its own, as well as in conjunction with the curvature drift effects. 
The results of this comprehensive analysis are presented in Figure~\ref{fig:dep_beta}. 
This reveals that the behaviour of the turbulent fluxes differs depending on the role played by $\beta_e$. 
With a consistently varying $\beta_e$ (the black line in Figure~\ref{fig:dep_beta}), there is a noticeable stabilizing impact on the fluxes. However, when the pressure gradient is held constant (orange line), the influence of $\beta_e$ appears to be minimal. 
Additionally, we note an escalation in transport (purple line) when drift velocities are proportionally increased with $\beta_e$, indicating that ETG modes are not significantly impacted by dynamical changes stemming from electromagnetic fluctuations. 
The $\beta_e$ scaling in our surrogate model, as described by Eq.~\eqref{eq:our_ETG_reduced_model}, should therefore primarily be interpreted as reflective of a pressure gradient scaling. 
This scaling corresponds to the effect of the Shafranov shift, for which $\beta_e$ serves as an effective substitute at fixed logarithmic gradients.
This interpretation is further corroborated by our finding that the electron heat flux is predominantly electrostatic. 
The electromagnetic contribution to the heat flux remains consistently minor and exhibits no discernible variation with changes in $\beta_e$ across the various scenarios we investigated.

\begin{figure}
\centering
\includegraphics[width=0.6\textwidth]{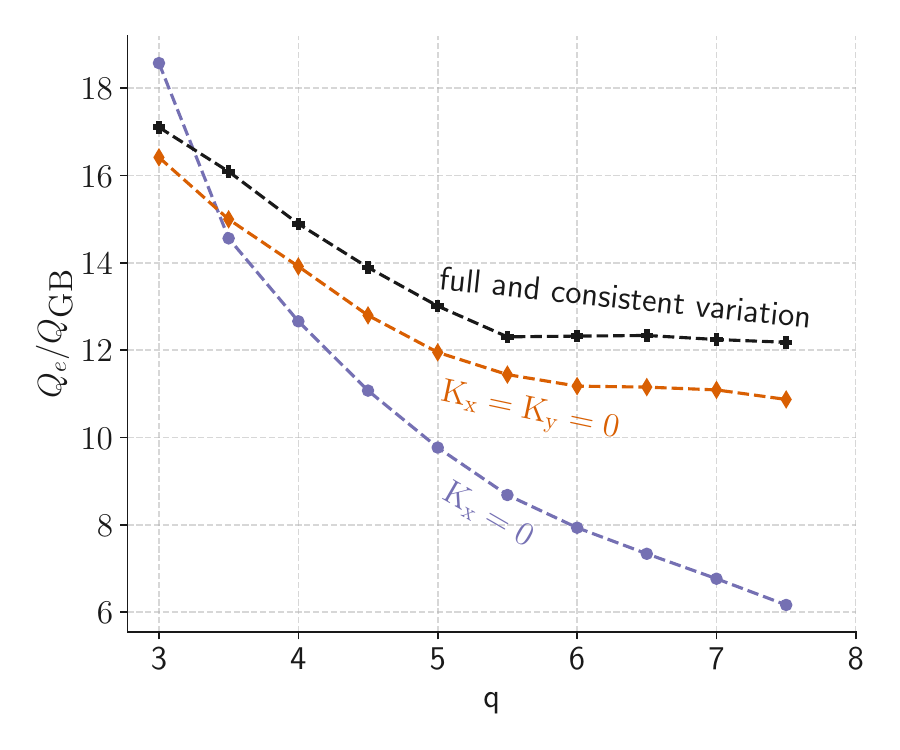}
\caption{Dependence of electron heat flux on $q$. The black line plots the results achieved when all geometric elements are calculated consistently. The orange line plots the results obtained when both $\rm K_x$ and $\rm K_y$ are artificially set to zero, and the purple line plots the results when only $\rm K_x$ is set to zero.}
\label{fig:dep_q}
\end{figure}

\begin{figure}
\centering
\includegraphics[width=\textwidth]{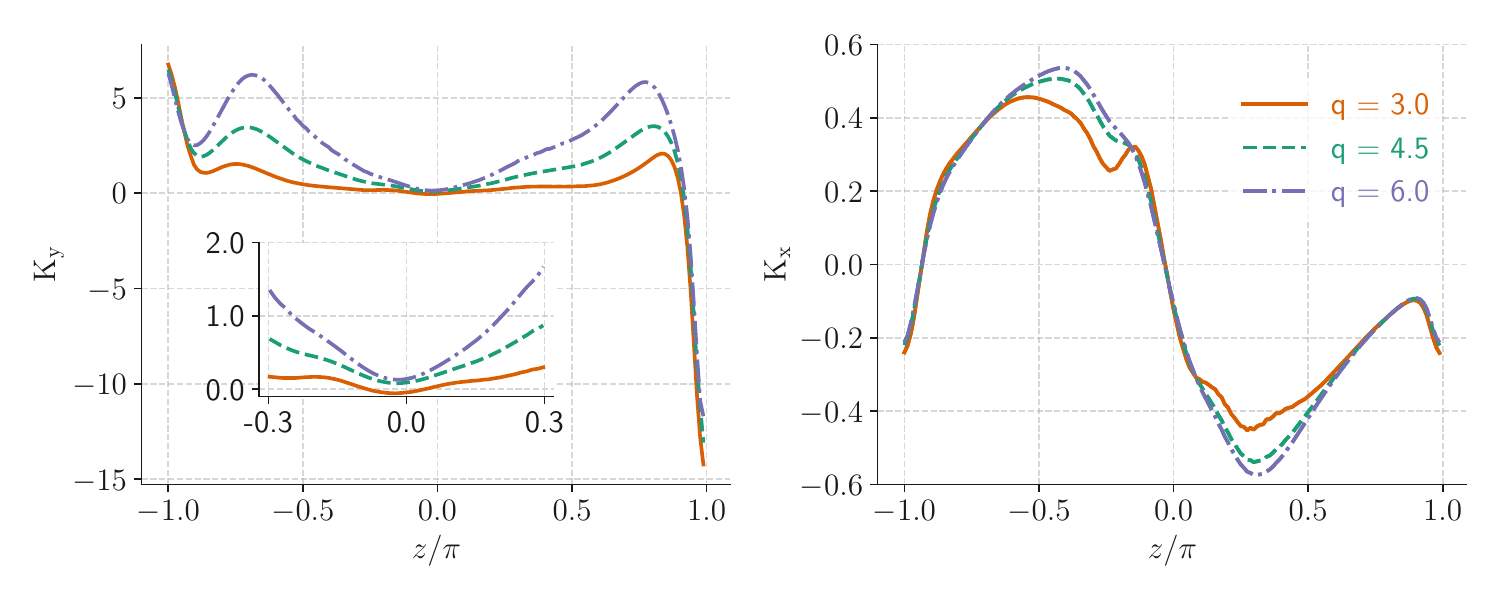}
\caption{Binormal $\rm K_y$ (left) and  radial $\rm K_x$ (right) components of the magnetic curvature as a function of the parallel coordinate $z$ for different values of the safety factor $q$. The inset in the left plot provides a detailed view of the range $-0.3 < z/\pi < 0.3$, demonstrating that at sufficiently large values of $q$, the curvature becomes strictly positive.}
\label{fig:curv}
\end{figure}

In our final analysis, we examine the impact of varying the safety factor $q$ beyond the range considered for training. 
It is important to note that while $q$ does not directly appear in the gyrokinetic equations, it is integral to the characterization of the magnetic geometry and therefore indirectly affects all related metric elements. 
The most significant influence arises from varying the magnetic curvature, as shown in Figure~\ref{fig:dep_q}.
Upon increasing the value of $q$, we observe a reduction of the turbulent fluxes, which eventually plateaus for $q \geq 6$. 
To better understand this pattern, we compare these results with instances where we purposefully set both the radial ($\rm K_x$) and binormal ($\rm K_y$) curvatures to zero. This particular case (orange line in Figure~\ref{fig:dep_q}) reveals a flux dependence similar to that in simulations preserving the complete geometry, although the levels of transport are marginally lower. 
Such a correlation upholds our hypothesis that the ETG modes in question tend to exhibit slab-like characteristics.
The influence of curvature on our system's behaviour is evident in the results where only $\rm K_y$ is considered (purple line in Figure~\ref{fig:dep_q}). 
Here, the fluxes decrease more rapidly with $q$, indicating that a higher safety factor causes a stabilizing modification of $\rm K_y$. 
This is confirmed in Figure~\ref{fig:curv}: for our specific case, the binormal curvature turns positive at all parallel positions once $q > 4$, balancing the destabilizing influence typically associated with $\rm K_x$.
Throughout all examined scenarios, including those where curvatures were artificially nullified, turbulence remains predominantly clustered around the outboard mid-plane. 
This consistency points to the prevalent role of finite Larmor radius effects in determining the parallel structure of the ETG modes, irrespective of curvature considerations.

\section{Conclusions} \label{sec:conclusions}

In the present work, we employed our recently developed sensitivity-driven dimension-adaptive sparse grid approximation method in the context of ETG turbulence in a tokamak pedestal, characterized by eight uncertain input parameters. 
This technique enabled us to determine an advanced surrogate model, which takes the form of the following scaling law:
\begin{equation*} 
    Q_e/Q_{\mathrm{GB}} = 0.038 \sqrt{\frac{m_e}{m_i}}\, \omega_{T_e}^{1.40} (\eta_e - 1)^{1.79} \tau_e^{-0.76} q^{-0.51} \beta_e^{-0.87} \lambda_{D}^{-0.51},
\end{equation*}
where $Q_{\rm GB}=n_eT_e^{5/2} m_i^{1/2}/(eB_0a)^2$.
To our knowledge, the identified dependencies on the safety factor $q$ and plasma beta $\beta_e$ are novel contributions, elucidating the influence of magnetic geometry on ETG modes. 
We further enhanced the robustness of our surrogate model by integrating a quantification of uncertainty in its predictions, which is of paramount importance in surrogate modelling. 
This was achieved through the computation of prediction intervals using the bootstrapping technique. 
Such measures are essential for evaluating the out-of-sample predictive accuracy of these models, extending their reliability beyond the scope of the training data.
We underpinned our approach with extensive numerical evidence, utilizing a total of $133$ test points -— including data points beyond the training set —- to demonstrate that our surrogate model delivers predictions with an acceptable level of precision.
The inclusion of prediction intervals not only adds rigour to our model's predictive capabilities but also affords a more profound understanding of its performance. 
Within the wider framework of turbulent transport simulations in fusion devices, our work implies that sensitivity-driven sparse grid approximations can be effectively harnessed to construct surrogate transport models directly from nonlinear, high-fidelity simulations, presenting a significant advancement in the field.

One application of the proposed surrogate model is profile reconstruction, which can be  done similarly to~\cite{Hatch_2024}.
Starting from an imposed boundary condition at the edge and an initial guess, the surrogate model can be used to provide an approximation of the heat flux that can be then fed to a transport solver evolving the electron profile iteratively until convergence.
Such an application is beyond our scope and is therefore left for future work.
In addition, further refinement of the surrogate model considering, for example, the value of the critical gradients as well as the impact of toroidal ETG is also left for our future research.

\section*{Acknowledgements}
This work was supported in part by the Exascale Computing Project (No.~17-SC-20-SC), a collaborative effort of the U.S.~Department of Energy Office of Science and the National Nuclear Security Administration.
Simulations were performed on the Leonardo supercomputer at CINECA, Italy.

\section*{Data availability statement}
The data that support the findings of this study are openly available in the repository at \url{https://github.com/ionutfarcas/surrogate_model_etg_pedestal}.


\appendix
\section{Set-up for the eight input plasma parameters}\label{app:params}

The set-up for the eight parameters describing the base ETG scenario is listed in table \ref{tab:parameters}.
This includes their respective nominal values (second column) and the corresponding left and right uniform bounds considered in the UQ and SA analysis in~\cite{FMJ22}.
We note that in~\cite{FMJ22}, the temperature and density gradients were normalized with respect to the major radius, $R$, and the output of interest was the electron heat flow computed in SI units (MW).

\begin{table}
\begin{center}
\def~{\hphantom{0}}
\begin{tabular}{cccc}
uncertain input parameter & nominal value & left uniform bound & right uniform bound \\
electron temperature $T_{e}~[\mathrm{keV}]$ & $0.3970$ & $0.3573$ & $0.4367$ \\
electron density $n_{e}$~[$10^{19} \mathrm{m^{-3}}$] & $4.4923$ & $4.0428$ & $4.9412$ \\
temperature gradient  $\omega_{T_e} = a/L_{T_e}$ & $39.9347$ & $31.9477$ & $47.9216$ \\
density gradient $\omega_{n_e} = a/L_{n_e}$ & $18.3928$ & $15.1150$ & $22.6726$\\
temperature ratio $\tau$ & $1.4400$ & $1.1520$ & $1.7280$ \\
effective ion charge $Z_\mathrm{eff}$ & $1.9900$ & $1.5920$  & $2.3880$\\
safety factor $q$ & $4.5362$ & $3.6289$ & $5.4434$ \\
magnetic shear $\hat{s}$ & $5.0212$ & $4.0169$ & $6.0254$ \\
\end{tabular}
\caption{Summary of the eight uniform uncertain parameters considered in~\cite{FMJ22}.
The second column shows their nominal (mean) value.
The corresponding left and right uniform bounds are listed respectively in the third and fourth columns.
The density and temperature gradients are normalized with respect to the minor radius, $a$.}
\label{tab:parameters}
 \end{center}
\end{table}

\section{Eight-dimensional sparse grid surrogate model}\label{app:SGI}
The sparse grid polynomial surrogate in GB units with $Q_{\rm GB}=n_eT_e^{5/2}m_i^{1/2}/(eB_0a)^2$, depending on all eight uncertain inputs listed in table~\ref{tab:parameters}, obtained from our UQ and SA study in~\cite{FMJ22}, expressed using a Legendre basis, reads:
\begin{equation} \label{eq:Q_SGI_8D_full}
\begin{split}
Q_{e}/Q_{\mathrm{GB}} = & 
0.19866664868934378 \tau \omega_{n_e} \omega_{T_e} - 1.1413477734069242 \tau^2 \omega_{n_e} + \\
& 1.1300923036977033 \tau^2 \omega_{T_e} + 0.012387562637082726 \omega_{n_e}^2 \omega_{T_e} - \\
& 0.08487445688736855 \tau \omega_{T_e}^2 + 0.009721433191939952q\omega_{T_e}^2 - \\
& 0.005966686182384836 \omega_{n_e} \omega_{T_e}^2 + 0.0014173787745147169 \omega_{T_e}^3 - \\ 
& 0.010912805921383906 \omega_{n_e}^3 - 0.14098572222663125 \tau \omega_{n_e}^2 + \\
& 0.0878050918300935q^2 \omega_{T_e} - 0.4779566243839404\omega_{n_e}^2T_e + \\
& 0.40612708864138414 \omega_{n_e}\omega_{T_e}T_e - 0.10152212813990472\omega_{T_e}^2T_e - \\
& 0.09610059439413138 q^3 - 6.565071206471347\tau^3 - \\ 
& 0.007095934720313662 \omega_{T_e}^2 n_e + 0.10660808146140052q\omega_{n_e} + \\
& 0.7209441445034814 \omega_{n_e}^2 + 2.3708835567864472 \tau \omega_{n_e} - \\ 
& 1.5507824853701202 q^2 - 0.6219694465716796\tau q - \\
& 1.5881501546239818 \tau\omega_{T_e} + 11.292418391099027 \tau^2 - \\ 
& 1.6038679243023q \omega_{T_e} + 0.37186341930335004 n_e^2 - \\ 
& 0.6593622299176898 \omega_{n_e}\omega_{T_e} + 0.16335509513250532 \omega_{T_e}^2 + \\ 
& 0.369546253431427 \omega_{T_e}n_e + 5.968742046874542 \tau T_e + \\ 
& 3.3170053895934304 q T_e + 7.075768404578394 \omega_{n_e} T_e - \\
& 2.5422978283907813 \omega_{T_e}T_e + 64.48667130778047 T_e^2 + \\
& 1.3682599362541525 T_e n_e + 0.2294389248146788 \omega_{n_e}n_e - \\ 
& 0.06963000365198978 \tau n_e + 0.3511888827695439 q n_e + \\
& 31.28952697970901 q - 10.12874179297306 \tau + \\
& 0.16860715589513 Z_{\textrm{eff}} - 9.035737970505384 \omega_{n_e} - \\
& 14.949102090605091 n_e + 0.19518454179911435 \hat{s} - \\
& 113.39196370792155T_e + 7.000425385025428 \omega_{T_e} - \\
& 11.083912068866233.
\end{split}
\end{equation}
It amounts to a multi-variate polynomial of third degree, comprising $47$ terms.

\section{Out-of-bounds testing data}~\label{app:out_ouf_bounds_testing_data}

Table~\ref{tab:out_of_distr_data} shows the values of the \textsc{Gene} parameters $\{\omega_{T_e}, \eta_e, q, \tau, \beta_e, \lambda_{D}, \hat{s}, Z_{\mathrm{eff}}\}$ (columns two to nine) as well as the corresponding electron heat fluxes in GB units (last column) for the $40$ out-of-bounds testing parameters used in Section~\ref{subsubsec:out_of_bounds_validation}.
\begin{table}
\centering
\begin{tabular}{cccccccccc}
index & $\omega_{T_e}$ & $\eta_e$ & $\tau$ & $q$ & $\beta_e$ & $\lambda_{D}=\lambda_{De}/\rho_e$ & $\hat{s}$ & $Z_{\mathrm{eff}}$ & $Q_e/Q_{\mathrm{GB}}$\\
1 & 33.0  & 3.0 & 0.55 & 4.0 & 0.0124 & 1.25707748 & 2.0 & 1.65 & 9.123 \\ 
2 & 31.0  & 2.0 & 0.775 & 3.5 & 0.01128 & 1.10252989 & 4.0 & 0.975 & 2.731 \\ 
3 & 26.0  & 4.0 & 0.325 & 4.5 & 0.01128 & 1.50249497 & 0.2 & 2.325 & 11.56 \\ 
4 & 10.62  & 2.5 & 0.6625 & 3.25 & 0.02606 & 1.04394603 & 1.0 & 1.988 & 0.4035 \\ 
5 & 30.62  & 3.5 & 0.8875 & 4.75 & 0.00784 & 1.1722308 & -1.0 & 1.312 & 10.0 \\ 
6 & 26.62  & 1.5 & 0.4375 & 3.75 & 0.00784 & 1.69504293 & 3.0 & 2.663 & 0.51 \\ 
7 & 17.28  & 1.75 & 0.6062 & 4.375 & 0.008979 & 1.82395972 & 4.5 & 1.819 & 0.4808 \\ 
8 & 25.53  & 4.75 & 0.8313 & 3.875 & 0.006004 & 1.42794407 & 2.5 & 1.144 & 17.4 \\ 
9 & 17.16  & 2.25 & 0.2688 & 4.125 & 0.008979 & 1.01795159 & 3.5 & 0.8063 & 2.869 \\ 
10 & 10.16  & 3.25 & 0.4938 & 3.625 & 0.02088 & 1.13577944 & 5.5 & 2.831 & 1.35 \\ 
11 & 30.35  & 2.625 & 0.3531 & 3.938 & 0.01428 & 1.46377777 & -0.25 & 0.3844 & 6.132 \\ 
12 & 11.85  & 4.625 & 0.8031 & 4.938 & 0.01596 & 1.08695585 & 3.75 & 1.734 & 0.825 \\ 
13 & 11.48  & 1.625 & 0.5781 & 3.438 & 0.01624 & 1.23413824 & 5.75 & 1.059 & 0.2112 \\ 
14 & 15.04  & 3.125 & 0.4656 & 4.688 & 0.008102 & 1.64002595 & 4.75 & 0.7219 & 3.852 \\ 
15 & 19.79  & 2.125 & 0.2406 & 3.188 & 0.008939 & 1.33435987 & 2.75 & 2.747 & 2.585 \\ 
16 & 45.66  & 4.375 & 0.4094 & 3.562 & 0.01817 & 1.05772043 & 1.25 & 0.5531 & 41.14 \\ 
17 & 46.16  & 2.375 & 0.8594 & 4.562 & 0.004406 & 1.39461942 & 5.25 & 1.903 & 13.07 \\ 
18 & 50.41  & 3.375 & 0.6344 & 3.062 & 0.008497 & 1.75599626 & 3.25 & 1.228 & 21.78 \\ 
19 & 31.73  & 3.875 & 0.9719 & 3.812 & 0.007288 & 1.54445643 & -1.75 & 2.241 & 10.76 \\ 
20 & 10.6  & 2.875 & 0.7469 & 4.312 & 0.01417 & 1.28134533 & 0.25 & 1.566 & 0.6607 \\ 
21 & 20.65  & 2.812 & 0.4797 & 3.594 & 0.007336 & 1.34869001 & -0.625 & 1.692 & 5.986 \\ 
22 & 10.84  & 3.812 & 0.2547 & 4.094 & 0.01115 & 1.66685386 & 1.375 & 1.017 & 4.193 \\ 
23 & 21.47  & 1.812 & 0.7047 & 3.094 & 0.01422 & 1.16278488 & 5.375 & 2.367 & 0.9622 \\ 
24 & 12.59  & 1.312 & 0.8172 & 3.844 & 0.01807 & 1.24544945 & 0.375 & 0.6797 & 0.1328 \\ 
25 & 32.59  & 2.312 & 0.1422 & 3.344 & 0.01008 & 1.48275739 & 2.375 & 1.355 & 6.441 \\ 
26 & 21.97  & 4.312 & 0.5922 & 4.344 & 0.008684 & 1.09465979 & -1.625 & 2.705 & 9.052 \\ 
27 & 12.83  & 2.062 & 0.9859 & 4.969 & 0.01587 & 1.29399103 & 5.875 & 0.5109 & 0.1904 \\ 
28 & 61.83  & 4.062 & 0.5359 & 3.969 & 0.01447 & 1.01174754 & 1.875 & 1.861 & 78.36 \\ 
29 & 32.83  & 3.062 & 0.7609 & 3.469 & 0.003641 & 1.56675126 & 3.875 & 2.536 & 21.71 \\ 
30 & 46.2  & 3.562 & 0.8734 & 3.719 & 0.007383 & 1.78897427 & -1.125 & 0.8484 & 17.02 \\ 
31 & 6.201  & 1.562 & 0.4234 & 4.719 & 0.00822 & 1.20199048 & 2.875 & 2.198 & 0.2105 \\ 
32 & 38.64  & 4.562 & 0.1984 & 3.219 & 0.02602 & 1.06480002 & 4.875 & 1.523 & 32.98 \\ 
33 & 44.76  & 2.562 & 0.6484 & 4.219 & 0.008917 & 1.4109689  & 0.875 & 2.873 & 11.04 \\ 
34 & 23.31  & 3.438 & 0.6766 & 4.406 & 0.01075 & 1.14457745 & 2.625 & 0.4266 & 6.115 \\ 
35 & 5.56  & 2.438 & 0.9016 & 3.906 & 0.01902 & 1.02427119 & 4.625 & 2.452 & 0.2475 \\ 
36 & 52.97  & 2.938 & 0.5641 & 3.156 & 0.009562 & 1.07942823 & 1.625 & 0.7641 & 34.79 \\ 
37 & 8.779  & 1.938 & 0.3391 & 3.656 & 0.004913 & 1.8610814 & 3.625 & 1.439 & 0.8462 \\ 
38 & 24.73  & 1.688 & 0.7328 & 4.031 & 0.005436 & 1.52304234 & 4.125 & 0.5953 & 1.088 \\ 
39 & 27.29  & 2.688 & 0.5078 & 4.531 & 0.0185 & 0.99968494 & -1.875 & 1.27 & 2.196 \\ 
40 & 14.26  & 4.188 & 0.6203 & 3.281 & 0.008179 & 1.37879224 & -0.875 & 0.9328 & 8.305 \\ 
\end{tabular}
\caption{Summary of the main parameters (columns two to nine, in the same units as in the main text) and of the corresponding GB-normalized electron heat flux (last column) for the $40$ out-of-bounds testing parameters used to validate our proposed surrogate model.}
\label{tab:out_of_distr_data}
\end{table}

\section{Unit conversion}~\label{app:coversion_to_Hatch_2024}
For simplicity, in the following we use superscript \emph{Fa22} to refer to quantities used in our original UQ study in~\cite{FMJ22} and \emph{Fa24} to refer to quantities used in the present paper.
Analogously, superscripts \emph{Ha22} and \emph{Ha24} are employed to refer to quantities specific to~\cite{Ha22} and~\cite{Hatch_2024}, respectively.

In~\cite{FMJ22}, our simulations adopted the generalized Miller geometry using $R = 1.68~\mathrm{m}$, with the temperature gradient defined as
\begin{equation} \label{eq:temp_grad_original}
\omega_{T_e}^{\mathrm{Fa22}} = \frac{R}{T_e}\frac{\mathrm{d}T_e}{\mathrm{d}r}.
\end{equation}
The conversion factor between $\omega_{T_e}^{\mathrm{Fa22}}$ in~\eqref{eq:temp_grad_original} and $\omega_{T_e}^{\mathrm{Fa24}}$ used in this paper is $4.6576$, therefore $\omega_{T_e}^{\mathrm{Fa24}} = \omega_{T_e}^{\mathrm{Fa22}}/4.5676$.

To use the model in~\eqref{eq:model_hatch_et_al_2024} in the present work, we must (i) convert the temperature gradient and Debye length in our units to be compatible with the units used in~\eqref{eq:model_hatch_et_al_2024} and (ii) convert the ensuing flux predictions to account for the area definition.
The Debye length conversion is straightforward and was explained in the main text.
To convert $\omega_{T_e}^{\mathrm{Fa24}}$ to $\omega_{T_e}^{\mathrm{Ha24}}$, we can use the chain rule to obtain
\begin{equation} \label{eq:temp_grad_conversion_H24}
\begin{split}
\omega_{T_e}^{\mathrm{Ha24}} = & \frac{1}{T_e}\frac{\mathrm{d} T_e}{\mathrm{d}\psi_N}=\frac{R}{T_e}\frac{\mathrm{d} T_e}{dr}\frac{1}{R}\frac{\mathrm{d} r}{d\psi_N}= \omega_{T_e}^{\mathrm{Fa22}} \frac{\psi_N}{R}\frac{\mathrm{d} r}{\mathrm{d} \psi} \\ = & 4.5676 \times \omega_{T_e}^{\mathrm{Fa24}} \frac{\psi_N}{R}\frac{\mathrm{d} r}{\mathrm{d} \psi}.
\end{split}
\end{equation}
Using the definition of the Miller geometry, we can evaluate $\mathrm{d} r/ \mathrm{d} \psi$ in~\eqref{eq:temp_grad_conversion_H24} for any set of input parameters.
The same is not true for $\psi_N/R$, which requires the knowledge of a global equilibrium. 
The value of this factor for our nominal parameters is 1/0.31, however since the results of ~\eqref{eq:model_hatch_et_al_2024} linearly scale with it, it is clear that a model comparison is inevitably skewed by the specific choice of this factor. 
Since the model in~\eqref{eq:model_hatch_et_al_2024} is richer than~\eqref{eq:model_hatch_et_al_2022}, we nevertheless employ it in Section~\ref{sec:results} for a more comprehensive overview using a constant value 1/0.20, which is close to the one for our nominal inputs.

The predictions $Q^{\mathrm{Ha24}}_{e}/Q^{\mathrm{Ha24}}_{\mathrm{GB}}$ provided by~\eqref{eq:model_hatch_et_al_2024} can be converted to $Q^{\mathrm{Fa24}}_{e}/Q^{\mathrm{Fa24}}_{\mathrm{GB}}$  as
\begin{equation*}
Q^{\mathrm{Fa24}}_{e}/Q^{\mathrm{Fa24}}_{\mathrm{GB}} = Q^{\mathrm{Ha24}}_{e}/Q^{\mathrm{Ha24}}_{\mathrm{GB}}~A^{\mathrm{Ha24}}/{V^{\prime}}^{\mathrm{Ha22}},
\end{equation*}
where $A^{\mathrm{Ha24}} = 51.07~\mathrm{m^2}$ and ${V^{\prime}}^{\mathrm{Ha22}} = 37.28 ~\mathrm{m^2}$.
Moreover, $Q^{\mathrm{Ha24}}_{\mathrm{GB}} = Q^{\mathrm{Fa24}}_{\mathrm{GB}} = Q^{\mathrm{Ha22}}_{\mathrm{GB}}$.


\bibliographystyle{jpp}
\bibliography{ETG_SG_surrogate_model}

\end{document}